\documentclass[runningheads]{llncs}
\usepackage{graphicx}
\usepackage[absolute,overlay]{textpos}
\usepackage{enumitem}
\usepackage{amsmath}
\usepackage{textcomp}
\usepackage{float}
\usepackage{xcolor}
\newcommand*{\figref}[1]{Fig.~(\ref{fig:#1})}

\begin{document}

\title{Reconstruction of radio signals from air-showers with autoencoder}

\author{
Pavel~Bezyazeekov\inst{1}
\and
Nikolay~Budnev\inst{1}
\and
Oleg~Fedorov\inst{1}
\and
Oleg~Gress\inst{1}
\and
Oleg~Grishin\inst{1}
\and
Andreas~Haungs\inst{2}
\and
Tim~Huege\inst{2,3}
\and
Yulia~Kazarina\inst{1}
\and
Matthias~Kleifges\inst{4}
\and
Dmitriy~Kostunin\inst{5}
\and
Elena~Korosteleva\inst{6}
\and
Leonid~Kuzmichev\inst{6}
\and
Vladimir~Lenok\inst{2}
\and
Nima~Lubsandorzhiev\inst{6}
\and
Stanislav~Malakhov\inst{1}
\and
Tatyana~Marshalkina\inst{1}
\and
Roman~Monkhoev\inst{1}\textbf{}
\and
Eleonora~Osipova\inst{6}
\and
Alexandr~Pakhorukov\inst{1}
\and
Leonid~Pankov\inst{1}
\and
Vasiliy~Prosin\inst{6}
\and
Frank~Schr\"oder\inst{2,7}
\and
Dmitriy~Shipilov\inst{1}
\and
Alexey~Zagorodnikov\inst{1}
}%
\authorrunning{P. Bezyazeekov et al.}

\institute{
	Institute of Applied Physics ISU, Irkutsk, Russia
	\and 
	Institut f\"ur Kernphysik, KIT, Karlsruhe, Germany
	\and 
	Astrophysical Institute, Vrije Universiteit Brussel, Pleinlaan 2, Brussels, Belgium
	\and
	Institut f\"ur Prozessdatenverarbeitung und Elektronik, KIT, Karlsruhe, Germany
	\and
	DESY, Zeuthen, Germany
	\and
	Skobeltsyn Institute of Nuclear Physics MSU, Moscow, Russia
	\and
	Bartol Research Inst., Dept. of Phys. and Astron., Univ. of Delaware, Newark, USA
}
\maketitle              
\begin{abstract}
The Tunka Radio Extension (Tunka-Rex) is a digital antenna array (63 antennas distributed over $\approx$~1 km$^{2}$) co-located with the TAIGA observatory in Eastern Siberia.
Tunka-Rex measures radio emission of air-showers induced by ultra-high energy cosmic rays in the frequency band of 30-80 MHz.
Air-shower signal is a short (tens of nanoseconds) broadband pulse.
Using time positions and amplitudes of these pulses, we reconstruct parameters of air showers and primary cosmic rays.
The amplitudes of low-energy event ($E<10^{17}$~eV) cannot be used for succesful reconstruction due to the domination of background.
To lower the energy threshold of the detection and increase the efficiency, we use autoencoder neural network which removes noise from the measured data.
This work describes our approach to denoising raw data and further reconstruction of air-shower parameters. We also present results of the low-energy events reconstruction with autoencoder.
\keywords{Tunka-Rex \and Efficiency \and Autoencoder \and Denoising}
\end{abstract}

\section{Introduction}

Cosmic rays (CR) are accelerated charged particles traveling in the space.
Most of them are protons, minor part is more massive atomic nuclei (up to iron).
The sources of CR are associated with stars at different evolution stages.
The ultra-high energy CR ($>10^{15}$ eV) carry information about most powerful cosmic accelerators and studying them is one of important tasks in modern astrophysics.
Due to the low flux of the ultra-high energy CR, it is impossible to measure them directly (in space or high layers of atmosphere), and they are detected by sparse ground detectors measuring cascades produced by their interaction with the atmosphere.
These cascades, called air-showers, consist of many secondary particles, including electrons and positrons, which produce short radio pulses due to deflection in the Earth's magnetic field and heterogenity of charge distribution in shower.  
These pulses have a broadband spectrum mostly in the MHz domain and a duration of tens of nanoseconds~\cite{Schroder:2016hrv}.

Tunka-Rex~\cite{Bezyazeekov:2015rpa} is a sparse antenna array located at the TAIGA facility~\cite{Budnev:2017fyg,Kostunin:2019nzy} in the Tunka Valley (Eastern Siberia).
It consists of 63 antennas measuring radio emission from air showers in the frequency band of 30-80 MHz.
Since Tunka-Rex is placed in a relatively radio-quiet location, the main background is from the Galaxy. 
However, there are plenty of non-stationary background sources, which  may distort the air-shower pulse and complicate the reconstruction of events with low energies.
In this work, we present our progress on the way of reconstruction of low-energy events by removing RFI from Tunka-Rex signal traces using the autoencoder (AE) neural network and discuss the performance of this approach.

\section{Dataset}

Tunka-Rex measures air-shower signals in two perpendicular polarizations and records it to traces of 1024 samples each with 200 MHz sampling.
For the reconstruction of cosmic-ray air-showers, the two main properties of radio pulses are used: the amplitude of the signal and its arrival time.
Details of this reconstruction are given in Refs.~\cite{Bezyazeekov:2015ica,Bezyazeekov:2018yjw}.

Before the reconstruction of the signals, we perform several preprocessing transformations.
Spectra of the signals obtained with the Fourier transform are cut by a digital bandpass to 35-75~MHz and filtered with a median filter, 
which removes narrow-band RFI and equalizes the noise using a sliding window of 3~MHz width.
Afterwards, the traces are upsampled in order to increase the timing resolution (factor 16 for this study).
Finally, the electric fields along the two polarization directions in the plane perpendicular to the shower axis are reconstructed, namely $\mathbf{v}\times \mathbf{B}$ (along the Lorentz force, where $\mathbf{v}$ is the direction of the air shower and $\mathbf{B}$ the direction of the geomagnetic field) and $\mathbf{v}\times\mathbf{v}\times\mathbf{B}$ perpendicular to it.
Since the main contribution of radio emission occurs in the $\mathbf{v}\times \mathbf{B}$ polarization, we consider only this one for the AE processing.

In this study, we use a dataset of 650\,000 samples of the measured background (2014-2017) recorded by Tunka-Rex and 25\,000 CoREAS~\cite{Huege:2013vt} simulations.
The air-shower pulse is randomly located within the signal window, summed with noise and folded with the Tunka-Rex hardware response taking into account the geometry of the air shower and the detector calibration.
As was discussed in Ref.~\cite{Bezyazeekov:2018yjw}, the simulated signals reproduce real ones with satisfactory accuracy.
As shown below, the methods developed for simulated pulses can be applied to the real data without additional tuning.

\section{Autoencoder (AE)}

\begin{figure}[t]
	\begin{center}
		\includegraphics[width=1.0\linewidth]{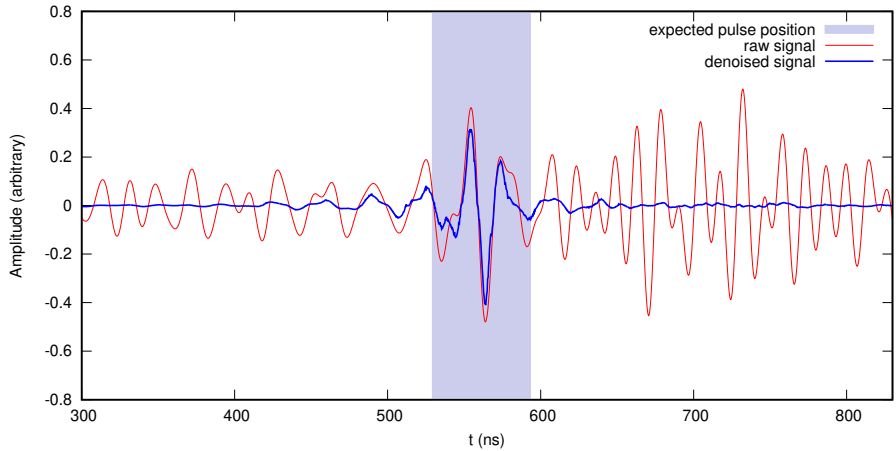}
		\caption{
			Example of the autoencoder performance on a measured Tunka-Rex trace showing successful denoising of the typical RFI after the signal.
		}
		\label{fig:real}
	\end{center}
\end{figure}

AE~\cite{Bengio:2013} is an unsupervised convolutional network used for learning the coded representation of the data and removing specific features from it.
The structure of AE can be described as follows: 
\begin{enumerate}
	\item \textbf{Encoder.} This part of AE distinguishes the features of noise contained in the input data by applying sets of filters.
	The filters perform the convolution of characteristic noise-related features with the input data, estimate its contribution as a result of the convolution, and afterwards send it to the max-pooling layer.
	The max-pooling layer performs a discrete downsampling of the data and sends it to the next convolution layer with the next set of filters.
	With each layer of the encoder, the data becomes more abstract and reduced in size.
	\item \textbf{Coded representation.} Central layer of AE has the least size (1024 in this study) and contains an abstract code of the input data.
	Due to its small size, we lost part of the input data.
	Learning procedure tunes AE for loosing noise-related data and leaving only cosmic-ray signals.
	\item \textbf{Decoder.} After encoding, the noise-related features are removed and the map of denoised data proceeds to the decoding part of AE, which produces a reverse reconstruction and returns a data array of the same dimension as the input.
	If successful, the resulting output is the denoised trace containing only the air-shower pulse, as shown in \figref{real}.
\end{enumerate}	 

\subsection{Configuration and training}

The input array for our AE consists of 4096 values, which corresponds to a trace length of 1280~ns and 0.3125~ns sampling in order to contain the signal window of 200~ns as well as surrounding background.
To minimize the loss, we normalize the input data to the [0:1] range with a baseline level at 0.5.

We have explicitly selected a subsample with low amplitudes and a low SNR for training to find out if the threshold may be lowered.
We implemented and trained our AE with Keras~\cite{chollet2015} and Tensorflow~\cite{tensorflow2015-whitepaper} in a uDocker container with GPU support.
After estimation of efficiency and accuracy of reconstruction with various depth (number of convolutional layers) and a number of filters per layer, we chose a 3-layer encoder with 8 filters per layer.
The full pipeline reconstruction using the data denoised by AE shows precision comparable with the standard method~\cite{bezyazeekov:2019dlc}.

\subsection{Real data reconstruction}

\begin{figure}[t]
	\begin{center}					
		\includegraphics[height=5.4cm]{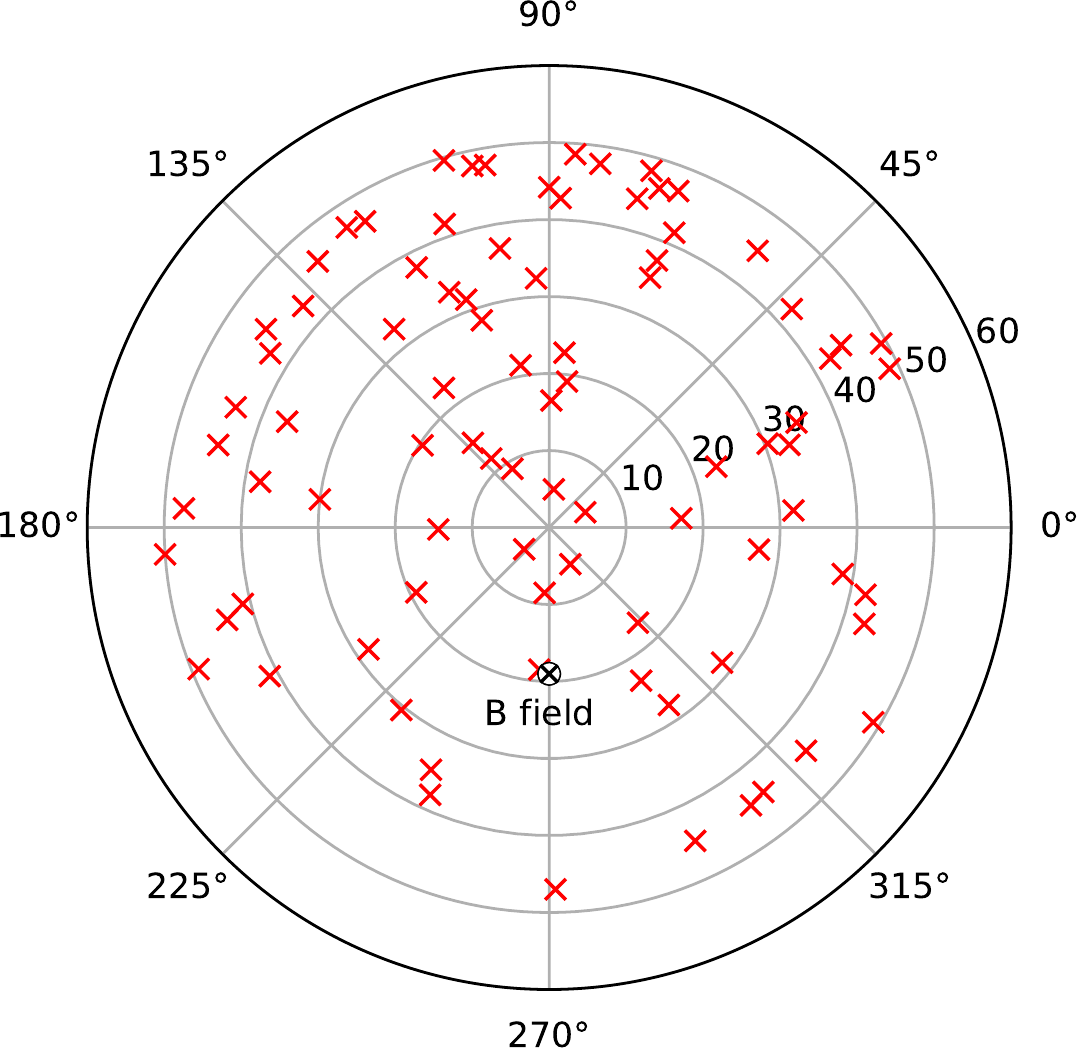}~~~~\includegraphics[height=5.4cm]{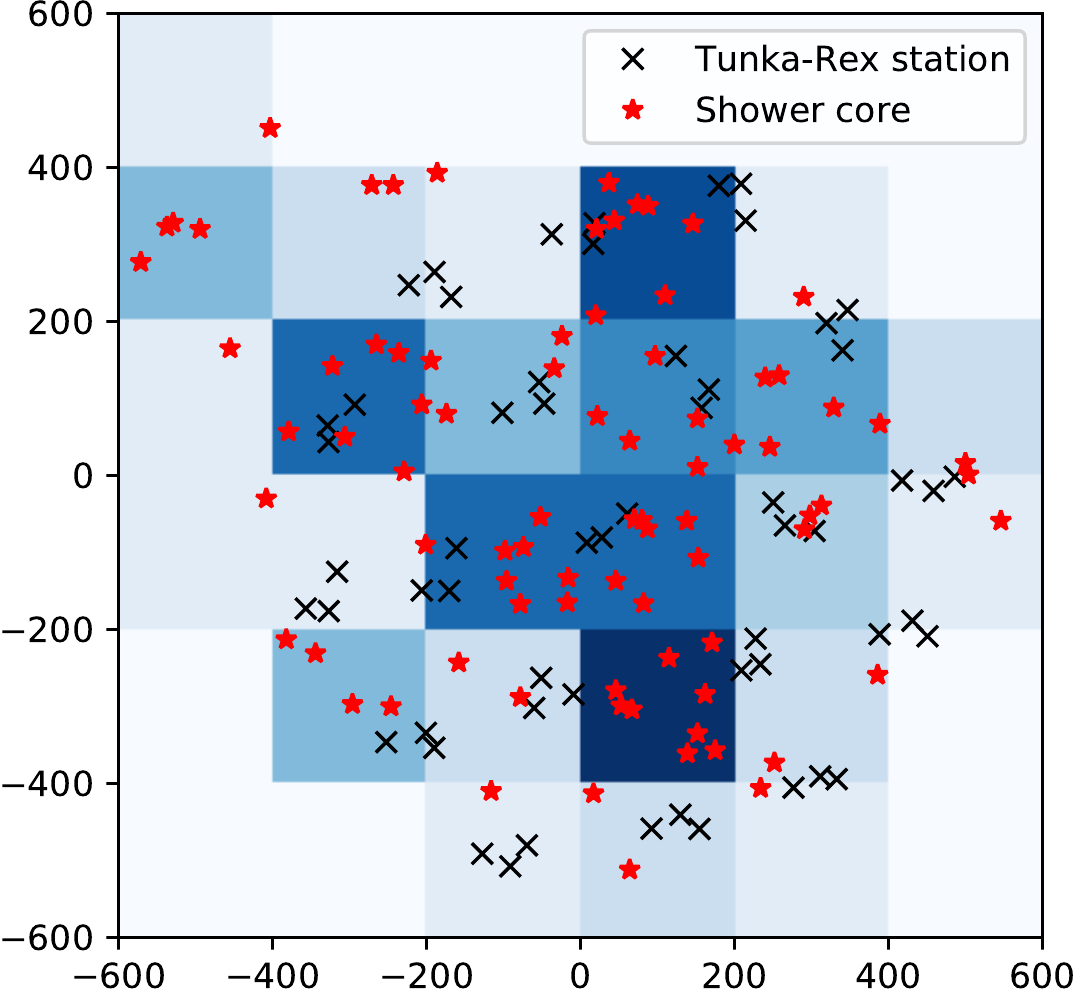}
		\caption{Left: Angular distribution of low-energy events used in this study. Right: distribution of shower cores over the surface of the detector.}
		\label{fig:distr}
	\end{center}
\end{figure}

\begin{figure}[t]
	\begin{center}
		Signal traces:\\
		\includegraphics[width=0.8\linewidth]{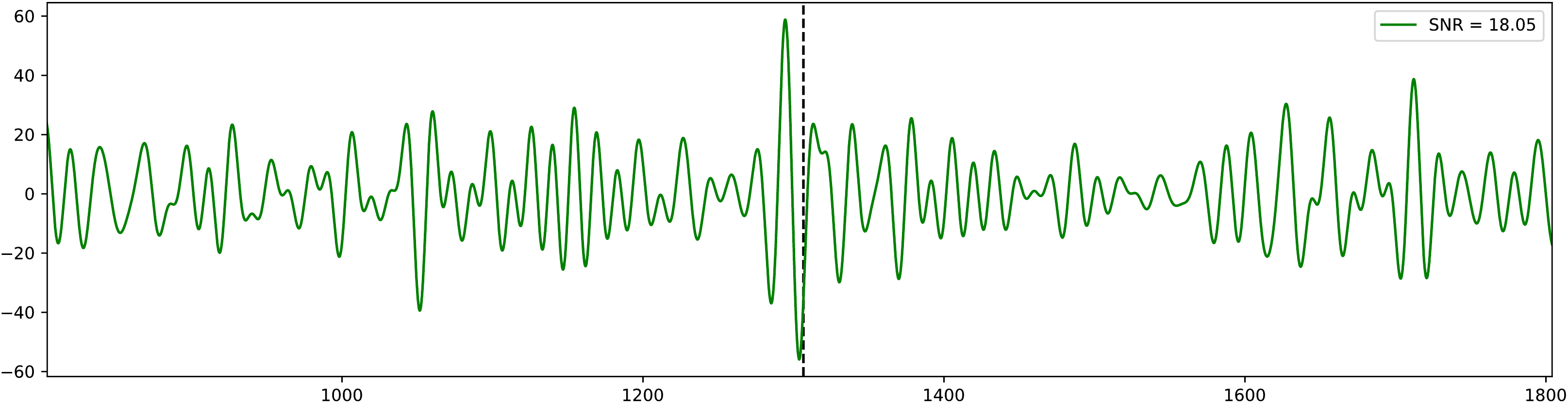}\\
		\includegraphics[width=0.8\linewidth]{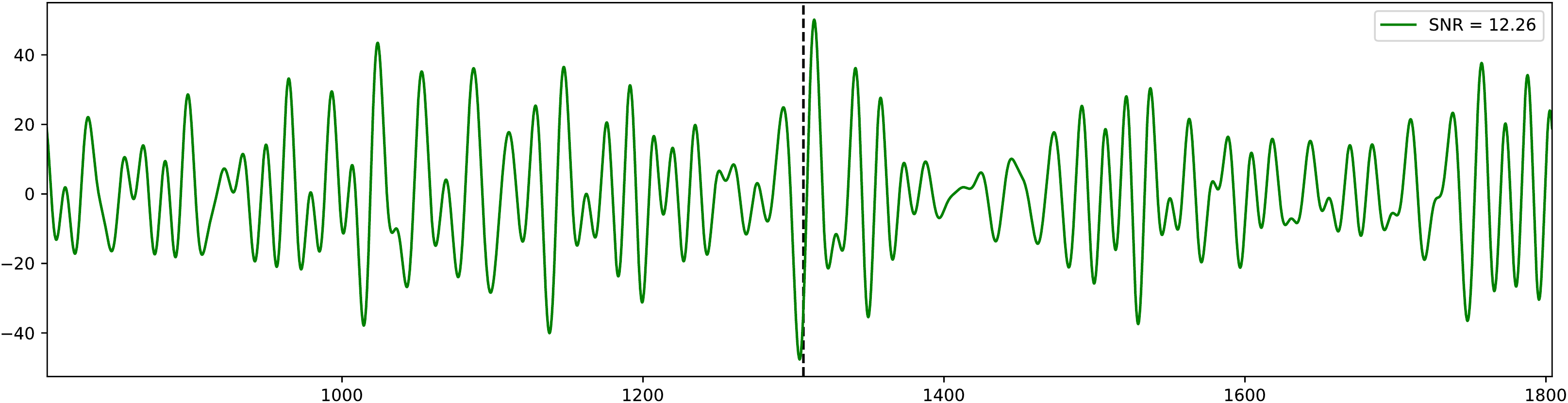}\\\includegraphics[width=0.8\linewidth]{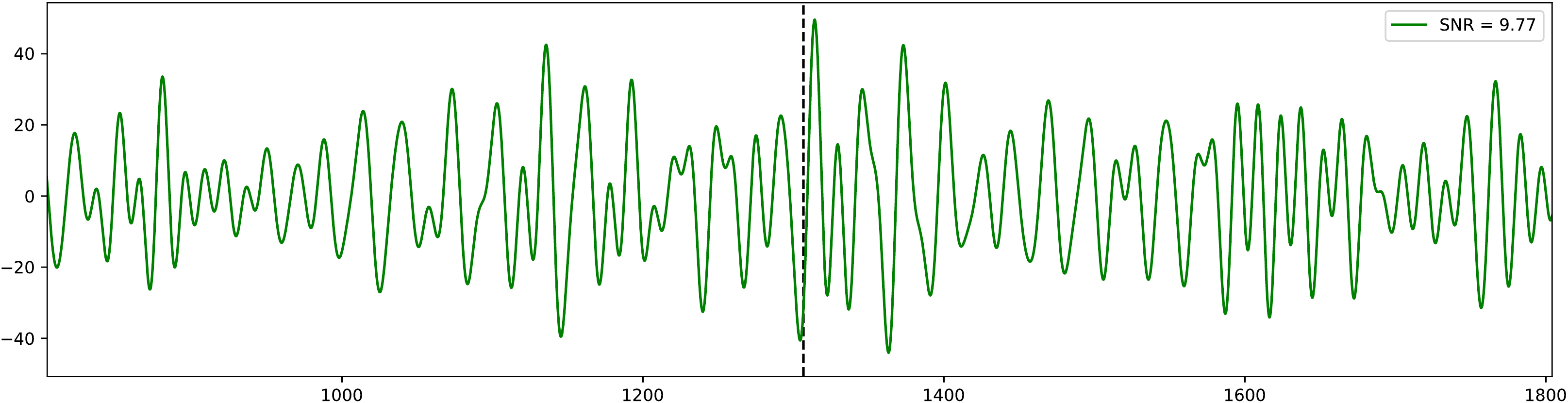}\\Coherent sum:\\
		\includegraphics[width=0.8\linewidth]{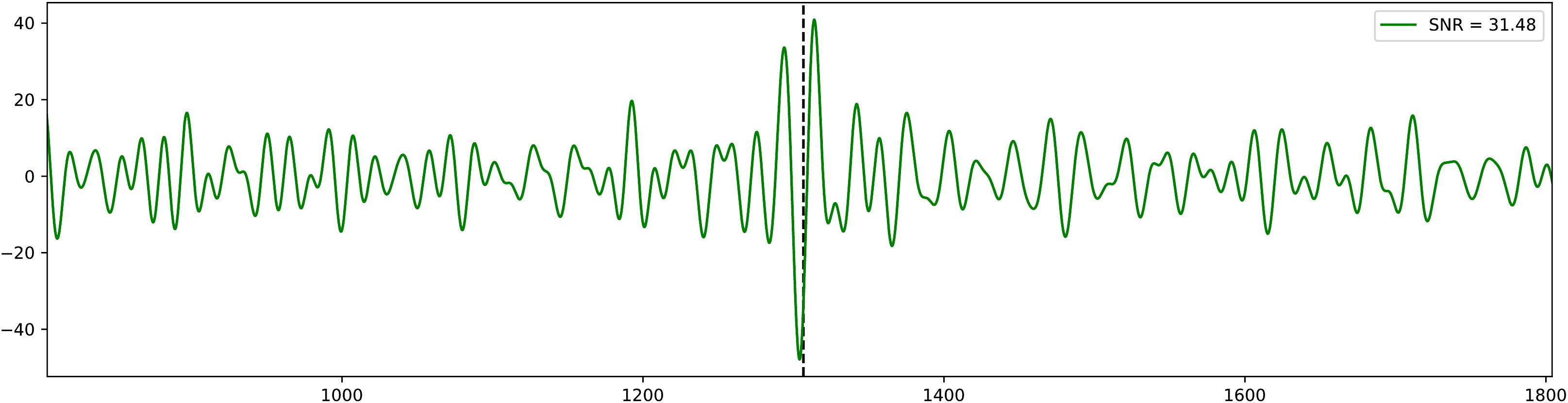}
		\caption{Example event with $E=30$~PeV. Top: radio traces recorded at different stations in an event. 
			Bottom: coherent sum of traces. 
			The dotted line shows the peak of the denoised trace appropriate to the air-shower signal timestamp.}
		\label{fig:summ}
	\end{center}
\end{figure}

After a series of tests on the simulated data, we test the performance of AE in application to the reconstruction of real low-energy events.
For this study, we use a set of low-energy Tunka-133~\cite{Prosin:2015voa} events~(\figref{distr}) with energies from 10$^{16}$ to 10$^{17}$~eV, which is unavailable for reconstruction using the standard Tunka-Rex method~\cite{Bezyazeekov:2015ica}.
The AE threshold was decreased from $0.395/0.500$ to $0.200/0.500$. The reconstruction pipeline is as follows:
\begin{enumerate}
	\item Traces (single polarization) in the event are processed with AE. 
	Peaks of envelopes of denoised traces are saved as assumed air-shower timestamps.
	\item Reconstruction of the shower front and arrival direction using these timestamps.
	\item Cut by applying the cross-check between the reconstruction direction by AE and by the host experiment Tunka-133: passing only events with the difference $
	<5^{\circ}$. 
	Additional cut for the geomagnetic angle $\alpha > 60^{\circ}$ to select events with the maximal contribution of the geomagnetic effect.
	\item Shifting the traces inside each event corresponding to the AE timestamps, summarizing them and normalizing to number of input traces in purpose of increasing SNR (\figref{summ}).
\end{enumerate}

\begin{figure}[t]
	\begin{center}
		\includegraphics[width=0.5\linewidth]{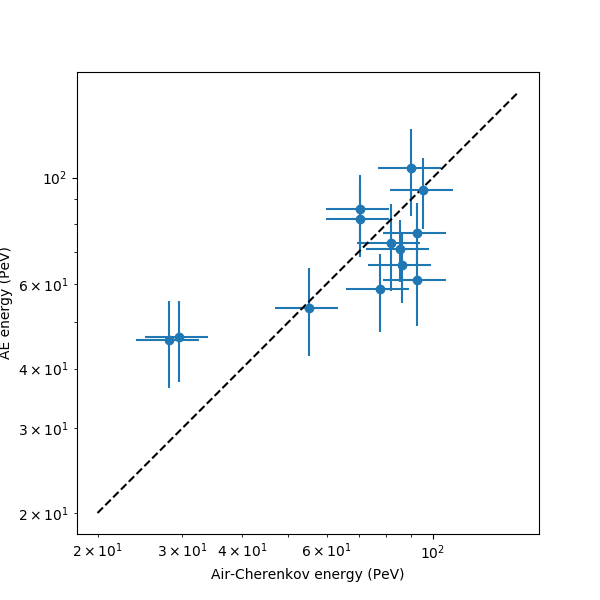}~\includegraphics[width=0.5\linewidth]{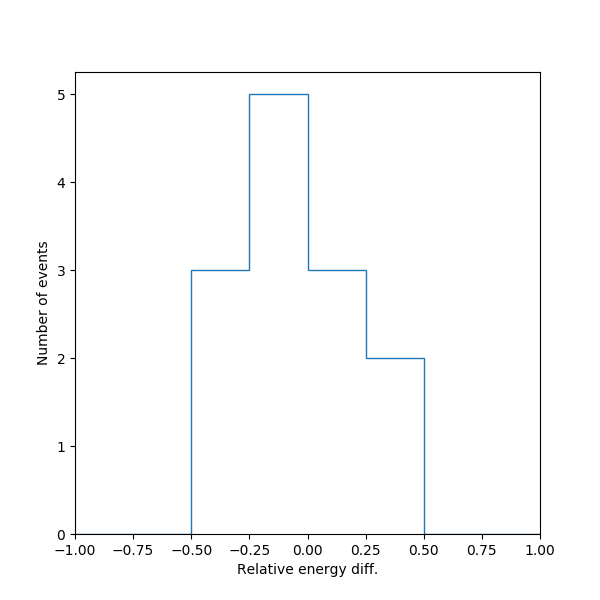}
		\caption{Left: AE energy vs Tunka-133 energy. Right: histogram of relative energy deviation.
		}
		\label{fig:energy}
	\end{center}
\end{figure}

The result of processing an event with this pipeline is an amplitude $S$ of the coherent sum at AE timestamps and a mean distance $r$ of input stations.
By this we extrapolate the lateral distribution function:
\begin{equation}
S_{0} = \frac{S}{\exp[\eta_{0}(r-r_{0})]},
\end{equation}
where $S_{0}$ is an amplitude at the distance $r_{0}$ from the shower axis, $\eta_{0} = -227.793\cdot10^{-5}$~m$^{-1}$ is correction factor.
This way we calculate the amplitude $S_{180}$ (180 m to axis) related to the best correlation with air-shower energy.
After that we reconstruct the energy $E$ using the single antenna method:
\begin{equation}
E = S_{180} \cdot \kappa,
\end{equation}
where $\kappa$ = $868\cdot10^{-6}\,\,\mathrm{EeV \cdot m/ V}$.
This way we reconstruct the set of low-energy events.
83 events passed the amplitude threshold and the arrival direction cross-check.
13 of them survived after $\alpha$ and SNR cuts.
In \figref{energy} one can see the results of this reconstruction.

\section{Discussion and conclusion}
The performance of Tunka-Rex AE has been tested on real data.
Reconstruction shows that we can reconstruct the arrival direction of low-energy events with AE, but energy precision for now is relatively low ($\approx$ 26\%).
We have illustrated the possibility of reconstruction of  low-energy events, and there is still room for improvement for the efficient application to the Tunka-Rex data processing.
We plan to improve AE by testing different loss metrics and network architectures with bigger dataset. Future work also implies modification of the input trace normalization  for saving the information of the absolute signal amplitude in denoised trace.
This will enable us to validate our technique on the Tunka-133 + Tunka-Rex data and check its performance in application to the data measured by Tunka-Rex + Tunka-Grande experiments.
In addition to the task of lowering the threshold, we also plan to check application of this technique to removing air-shower pulses from the raw data flow within the frame of the Tunka-21cm experiment~\cite{Kostunin:2019gho}.

\section*{Acknowledgements}
The work of P.Bezyazeekov on section "Real data reconstruction" is supported by the Russian Foundation for Basic Research "Mobility" program grant 19-32-50147. 
The work has been supported by Russian Federation for Basic Research grant 18-32-20220, the Helmholtz grant HRSF-0027, the Russian Federation Ministry of Science and High
Education (project. FZZE-2020-0024), the Mathematical Center in Akademgorodok under agreement No 075-2019-1675 with the Ministry of Science and Higher Education of the Russian Federation and Irkutsk State University grant 091-19-213.
P.Bezyazeekov thanks the community of the Institute for Nuclear Research, where this study has been carried out, and personally G. Rubtsov.
\bibliographystyle{ieeetr}
\bibliography{references}

\end{document}